\documentclass[12pt]{article}

\usepackage{graphicx}
\usepackage{amsfonts}
\usepackage{amsthm} 
\usepackage{amsmath}
\usepackage[top=3cm, bottom=3.25cm, left=2.75cm, right=2.75cm]{geometry}

\newtheorem*{theorem}{Theorem}

\newcommand{\Section}[1]{\section{#1} \setcounter{equation}{0}}
\newcommand{\be}{\begin{eqnarray}}
\newcommand{\ee}{\end{eqnarray}}
\newcommand{\beq}{\begin{equation}}
\newcommand{\eeq}[1]{\label{#1}\end{equation}}
\newcommand{\ber}{\begin{eqnarray}}
\newcommand{\eer}[1]{\label{#1}\end{eqnarray}}

\def\+{{+\!\!\!+}}

\newcommand{\nn}{\nonumber}
\newcommand{\kah}{K\"ahler~}

\begin{document}
\renewcommand{\theequation}{\thesection.\arabic{equation}}
\setcounter{page}{0}
\thispagestyle{empty}

\begin{flushright} \small
UUITP-12/10  \\  YITP-SB-10-10\\ Imperial-TP-2010-CH-02\\
\end{flushright}
\smallskip
\begin{center} \LARGE
{\bf Generalized Calabi-Yau metric and \\  
Generalized Monge-Amp\`ere equation}
\\[12mm] \normalsize
{\bf   Chris M. Hull$^{a}$, Ulf~Lindstr\"om$^{b}$, Martin Ro\v cek$^{c}$, \\
Rikard von Unge$^{d}$ and Maxim Zabzine$^{b}$} \\[8mm]
{\small\it
$^a$ The Blackett Laboratory, Imperial College London\\
Prince Consort Road, London SW7 2AZ, U.K.\\
~\\
 $^b$Theoretical Physics, Department of Physics and Astronomy,\\
Uppsala University,  Box 516,
SE-751 20 Uppsala, Sweden \\
~\\
$^c$C.N.Yang Institute for Theoretical Physics, Stony Brook University, \\
Stony Brook, NY 11794-3840,USA\\
~\\
$^{d}$Institute for Theoretical Physics, Masaryk University, \\
611 37 Brno, Czech Republic \\~\\}
\end{center}
\vspace{10mm}
\centerline{\bf\large Abstract}
\bigskip
\noindent
In the neighborhood of a regular point, generalized \kah geometry admits a
description in terms of a single real function, the generalized \kah potential.
We study the local conditions for a generalized \kah manifold to be a generalized 
Calabi-Yau manifold and we derive a non-linear PDE that the generalized \kah
potential has to satisfy for this to be true.
This non-linear PDE can be understood as a generalization of the complex
Monge-Amp\`ere equation and its solutions give supergravity solutions
with metric, dilaton and $H$-field. 
\eject
\footnotesize
\normalsize
\eject

 \Section{Introduction}
 
Generalized geometry, initiated by N.~Hitchin in \cite{hitchin} has 
attracted considerable interest in both physics and mathematics.
The essential idea behind generalized geometry is to replace the tangent bundle
$T$ by $T\oplus T^*$, the tangent plus the cotangent bundle. 
This replacement leads to many new and interesting 
geometrical concepts which can  play a prominent role in string
theory. The subject was further developed by Hitchin's
students \cite{gualtieriPhD, Cavalcanti:2005hq, Witt:2005sk}. 
   
In this paper we study the conditions for a generalized \kah manifold
to admit a generalized Calabi-Yau metric structure.
Our analysis is local and we do not consider global 
issues. We take as a starting point our previous results \cite{Lindstrom:2005zr} on
the local 
description of generalized \kah manifolds in terms of a single function, the generalized 
\kah potential. We derive a non-linear PDE for this potential 
which comes from the generalized Calabi-Yau conditions.  We will refer to this non-
linear PDE as the generalized Monge-Amp\`ere equation, since it generalizes the
complex Monge-Amp\`ere equation which appears in the local description of the
standard Calabi-Yau manifolds. This idea has  previously been discussed in
\cite{Halmagyi:2007ft} and some partial results were presented there. 
Here we give a full derivation for the general case. 
  
On the physics side we relate the Type II supergravity solutions for metric,
dilaton and NS-flux to the world-sheet description in terms of the $N=(2,2)$
non-linear sigma model. The generalized Calabi-Yau conditions appear naturally in 
supergravity solutions, \cite{Grana:2004bg}, \cite{Jeschek:2004wy} and
\cite{Grana:2005sn} (see \cite{Witt:2009bt} for a nice summary and further references).
On the other hand, generalized \kah manifolds \cite{gualtieriPhD} of any complex dimension appear as target
spaces for sigma models with $N=(2,2)$ supersymmetry \cite{Gates:1984nk} and
classically any generalized \kah potential $K$ (satisfying some minor conditions
required for the geometry to be non degenerate) gives a generalized \kah manifold.
The proposed generalized Monge-Amp\`ere equation appears at the quantum
level when one requires conformality of the sigma model at the one loop level.
The connection to standard 10 dimensional supergravity solutions come when
one also imposes the correct value of the central charge.
 
The paper is organized as follows: In Section \ref{GK-CY} we review the basic notions of 
generalized complex, generalized \kah and generalized Calabi-Yau geometries. Then,
in Section \ref{G-MA}, using the local description of generalized \kah geometry,
we present explicit formulas for the pure spinors that encode the geometry
in terms of the generalized \kah potential.  We use the fact that the generalized 
Calabi-Yau metric structure 
is equivalent to the existence of supersymmetric
supergravity solutions\footnote{That is, solutions that preserve some of the
supersymmetry of the target.}. We show that the compatibility conditions of the pure 
spinors
are trivially satisfied except for the normalization condition which gives a
non-linear PDE (the generalized Monge-Amp\`ere equation) 
for the generalized \kah potential.
Section \ref{special-cases} deals with a few special cases of the generalized
Monge-Amp\`ere equation and some simple solutions. 
In Section \ref{geometry} we elaborate on the geometrical aspects of the generalized 
Monge-Amp\`ere equation.
In Section \ref{physics} we comment on the relation of our results to the $\beta$-function 
calculations for supersymmetric sigma model. Section \ref{summary} gives the summary 
and discusses some open questions.
 
\section{Generalized \kah and Calabi-Yau geometries}
\label{GK-CY}
 
In this section we briefly  review the relevant concepts in generalized geometry.
Since our analysis is local we ignore complications related to global issues\footnote
{Concretely, we work in a contractible coordinate patch and
assume that $H = dB$ is exact in that patch, so that the effect of H can be
encoded into the appropriate B-transform. We do not need to use the twisted Courant 
bracket and the twisted de Rham differential here, and we do not discuss global issues.}.
This section does not contain  original material. For further details the reader may 
consult \cite{gualtieriPhD, Witt:2009bt, gualtieri2}. 
  
Consider a smooth manifold $M$. The tangent plus cotangent bundle $T\oplus T^*$
has a natural inner product
   \be
   \label{T+Tstarmetric}
    \langle X + \eta, Y + \xi \rangle = \frac{1}{2} (i_X \xi + i_Y \eta)~,
   \ee
where  $X$, $Y$ are vector fields and $\eta$, $\xi$ are one-forms.
The Courant bracket on $T\oplus T^*$ is 
\be
       [ X+ \eta , Y + \xi ]_c = \{ X, Y\} + {\cal L}_X \xi - {\cal L}_Y \eta - \frac{1}{2} d (i_X 
       \xi - i_Y \eta)~, 
\ee
and is a generalization of the Lie bracket naturally defined on $T$. A {\em generalized
complex structure} is defined as a decomposition
$(T\oplus T^*) \otimes \mathbb{C} = L + \bar{L}$ where $L$ is a maximally isotropic 
sub-bundle involutive with respect to the Courant bracket and 
$\bar{L}$ is its complex conjugate. 
This decomposition can be encoded in terms of an endomorphism ${\cal J}$ of
$T\oplus T^*$ with some additional
properties, such that $L$ is the  $+i$-eigenbundle of ${\cal J}$.

A {\em generalized \kah structure} is defined as two commuting generalized complex 
structures ${\cal J}_1$ and ${\cal J}_2$ such that the quadratic form $\langle {\cal J}_1 
{\cal J}_2 (X+ \xi), (X+\xi) \rangle $ is positive definite.  Equivalently
the generalized \kah structure gives rise to the decomposition          
$(T \oplus T^*) \otimes \mathbb{C} = L_1^+ \oplus L_1^- \oplus \bar{L} _1^+ \oplus 
\bar{L}_1^-$ where $L_1^+$ is the $+i$-eigenbundle 
for both of ${\cal J}_{1,2}$ and $L_1^-$ is the $+i$-eigenbundle for ${\cal J}_1$
and $-i$-
eigenbundle for ${\cal J}_2$. 
Due to Gualtieri's theorem \cite{gualtieriPhD}, generalized \kah geometry is equivalent to 
bihermitian Gates-Hull-Ro\v{c}ek geometry \cite{Gates:1984nk} which appears as the 
target space geometry for $N=(2,2)$ supersymmetric sigma models.
We review the details of Gates-Hull-Ro\v{c}ek geometry and the Gualtieri map in the 
appendices.

Differential forms play a crucial role in generalized geometry. We may regard the 
differential forms on $M$ as spinors for the bundle $T\oplus T^*$. A section
$X+\xi$ of $T+ T^*$ acts on a form $\rho$ according to
\be
   (X + \xi) \cdot \rho = i_X \rho + \xi \wedge \rho
\ee
and this satisfies the Clifford algebra identity for the indefinite metric (\ref{T+Tstarmetric}) 
on $T\oplus T^*$. An invariant bilinear form on spinors (differential forms) is given by 
\be
    \label{mukaipairing}
    (\rho_1, \rho_2) = \sum\limits_j (-1)^j [\rho_1^{2j} \wedge \rho_2^{d-2j} + \rho_1^{2j
    +1} \wedge 
    \rho_2^{d-2j-1}]~,
\ee
where $\rho_1 = \sum \rho_1^i$ and $\rho_2 = \sum \rho_2^i$ are linear
combinations forms of different degrees, the upper index indicating the form degree.
This is called the Mukai pairing of forms.
The spinor is called pure if it annihilates a maximal isotropic subspace of $T\oplus T^*$
(or its complexification $(T\oplus T^*)\otimes \mathbb{C}$).

A {\em generalized 
Calabi-Yau structure} is defined as a closed pure spinor $\rho$ such that 
$(\rho, \bar{\rho}) \neq 0$ everywhere on $M$, where $\bar{\rho}$ is the complex
conjugate of $\rho$. A closed pure spinor $\rho$ gives rise to a decomposition
$(T\oplus T^*) \otimes \mathbb{C} = L + \bar{L}$ where $L$ is a maximally isotropic
sub-bundle involutive with respect to the Courant bracket. Therefore a generalized
Calabi-Yau manifold is a special case of a generalized complex manifold.

A {\em generalized Calabi-Yau metric structure}\footnote{There is no fully accepted
use of term ``generalized Calabi-Yau'' and different authors use different adjectives like 
``weak'', ``strong'' to represent different version of the generalized Calabi-Yau condition. 
We follow Gualtieri's terminology \cite{gualtieriPhD}.} is defined as a pair of closed
pure spinors $\rho_1$ and $\rho_2$ such that the corresponding generalized complex 
structures ${\cal J}_1$ and ${\cal J}_2$ give rise to a generalized \kah structure and 
moreover that $(\rho_1, \bar{\rho}_1) = \alpha (\rho_2, \bar{\rho}_2)\neq 0$ for some
non-zero constant $\alpha$. 
       
The geometric data encoded by the pair of pure spinors defining a generalized
Calabi-Yau metric structure may be found by observing that it gives rise to 
generalized \kah geometry. Using the Gualtieri map we may then read off the metric
$g_{\mu\nu}$ and the closed NS 3-form $H_{\mu\nu\rho}$. 
Furthermore the dilaton $\Phi$ is defined via normalization of pure spinors as follows
\be
      \label{twospinorsCYrel}
       (\rho_1,\bar{\rho}_1) = \alpha (\rho_2,\bar{\rho}_2)= 
       e^{-2\Phi} {\rm vol_g} =
       e^{-2\Phi}\sqrt{g}~dx^1\wedge\ldots\wedge dx^D,
\ee
where $g= \det (g_{\mu\nu})$.  This data $(g_{\mu\nu}, H_{\mu\nu\rho}, \Phi)$ coming 
from the generalized Calabi-Yau metric structure 
is a Type II supersymmetric supergravity solution. It automatically solves the 
equation
\be
 \label{oneloopeqsmotion}
R_{\mu\nu}^{(+)} + 2 \nabla^{(-)}_\mu \partial_\nu \Phi = 0~,
\ee
where  $R^{(\pm)}_{\mu\nu}$ is the Ricci tensor for the connection with torsion
$\nabla^{(\pm)} = \nabla \pm \frac{1}{2}g^{-1} H$
where $\nabla$ is the Levi-Civita connection. 
Indeed, the symmetric and antisymmetric parts of (\ref{oneloopeqsmotion}) are the 
bosonic equations of motion of Type II supergravity with the RR-fields set to zero.  
For further relevant results and references the reader may consult the nice review
\cite{Witt:2009bt}. 

Our strategy will be to take our local description of generalized \kah geometry in
terms of the potential $K$ and construct the corresponding closed pure spinors for two 
generalized complex structures. Through the theorem in 
\cite{Jeschek:2004wy, Witt:2006tq} we know that the existence
of a pair of compatible closed pure spinors implies the existence of supersymmetric 
supergravity solutions. Imposing the generalized Calabi-Yau metric conditions (i.e.
the compatibility conditions for the pure spinors) we arrive at a non-linear PDE for
$K$. The Gualtieri map plays a crucial role in our 
construction.

\section{Generalized Monge-Amp\`ere equation}
\label{G-MA}

Recently, in \cite{Lindstrom:2005zr}, it was shown that, in the neighborhood 
of a regular point, generalized \kah geometry can be encoded 
in terms of a single real function: the generalized \kah potential $K$. The construction 
roughly goes as follows. We take ${\mathbb C}^{\frac{D}{2}}$ and divide the complex 
coordinates in four groups ${\mathbb C}^{\frac{D}{2}} = {\mathbb C}^{d_c} + 
{\mathbb C}^{d_t} +  {\mathbb C}^{d_s} + {\mathbb C}^{d_s}$
with the following notations for the coordinates
$\phi = (\boldsymbol{\phi}, \bar{\boldsymbol{\phi}})$,  $\chi = (\boldsymbol{\chi}, \bar
{\boldsymbol{\chi}})$, $X_L = (\boldsymbol{X}_l, \bar{\boldsymbol{X}}_{\bar{l}})$, 
$X_R = (\boldsymbol{X}_r, \bar{\boldsymbol{X}}_{\bar{r}})$ 
in the respective subspaces. 
The coordinates are adapted to the complex structures $J_\pm$ but in a more involved
way than in ordinary \kah geometry. In particular $\boldsymbol{\phi},
\boldsymbol{\chi}$ and $\boldsymbol{X}_l$ are holomorphic with respect to $J_+$
while there is no such simple relation between $X_R$ and $J_+$. On the other
hand, $\boldsymbol{\phi},\bar{\boldsymbol{\chi}}$ and $\boldsymbol{X}_r$ are 
holomorphic with respect to $J_-$ while there is no such simple relation between
$X_L$ and $J_-$.

Introducing the real function $K (\phi, \chi, X_L, X_R)$ which we call the generalized
\kah potential we derive formulas for $(g, J_\pm, H)$ in terms of derivatives
of $K$ such that they become the geometric data for a generalized \kah
manifold. In appendix \ref{appendix1} we review the relevant relations. 

In the language of supersymmetric sigma models, the function $K$ corresponds 
to a Lagrangian density in $N=(2,2)$ superspace and the coordinates correspond
to different sets of superfields: $(\boldsymbol{\phi}, \bar{\boldsymbol{\phi}})$ for 
chiral and anti-chiral; $\chi = (\boldsymbol{\chi}, \bar{\boldsymbol{\chi}})$ for twisted 
chiral and twisted anti-chiral; $X_L = (\boldsymbol{X}_l, \bar{\boldsymbol{X}}_{\bar{l}})$ 
for left semi-chiral and left semi-anti-chiral; $X_R = (\boldsymbol{X}_r,
\bar{\boldsymbol{X}}_{\bar{r}})$ for right semi-chiral and right semi-anti-chiral.
The sigma model provides a useful way of deriving and manipulating  the rather 
complicated expressions for the various geometrical objects. 
For further details about the sigma model interpretation the reader may consult
\cite{Lindstrom:2005zr, Lindstrom:2007qf, Lindstrom:2007xv}. 

It is important to stress that the choice of generalized \kah potential is not unique.
There are many choices that lead to the same geometric objects. This ambiguity
in the choice of $K$ can be understood both 
from geometry and from the sigma model point of view, see \cite{Hull:2008vw}. 

In the following discussion we adopt the following short-hand notations for the 
derivatives of $K$:
$K_C = \partial_\phi K = (K_c, K_{\bar{c}}) =( \partial_{\boldsymbol{\phi}} K, \partial_{\bar
{\boldsymbol{\phi}}} K) $, $K_T = 
\partial_\chi K = (K_t, K_{\bar{t}})= (\partial_{\boldsymbol{\chi}} K, \partial_{\bar
{\boldsymbol{\chi}}} K)$,
$K_L = \partial_{X_L} K = (K_l, K_{\bar{l}}) = ( \partial_{\boldsymbol{X}_l} K, \partial_{\bar
{\boldsymbol{X}}_{\bar{l}}}K)$,
$K_R = \partial_{X_R} K = (K_r, K_{\bar{r}}) = ( \partial_{\boldsymbol{X}_r} K, \partial_{\bar
{\boldsymbol{X}}_{\bar{r}}}K)$,
where we suppress all coordinates indices.  Analogously we define the matrices of
double derivatives of $K$, e.g. $K_{l\bar{r}}$ is our notation for the matrix of 
second derivatives
$\partial_{\boldsymbol{X}_l} \partial_{\bar{\boldsymbol{X}}_{\bar{r}}} K$ etc. 
We use matrix notation and suppress all indices.  

Next we would like to use $K$ to construct the pure spinors which encode the
generalized \kah geometry. Our ansatz for the pure spinors is 
\be
\label{definitionpurespinor}
\rho_{1,2} = N_{1,2}\wedge e^{R_{1,2}+iS_{1,2}}~,
\ee
 with the following notation
\be
\nn N_1 &=&  e^{f(\boldsymbol{\phi})}~d\boldsymbol{\phi}^1\wedge\ldots\wedge d
\boldsymbol{\phi}^{d_c}  ~, \\
\nn N_2 &=&  e^{g(\boldsymbol{\chi})}~d\boldsymbol{\chi}^1\wedge\ldots\wedge d
\boldsymbol{\chi}_{d_t} ~,\\
\nn R_1 &=& -d(K_L dX_L)~,\\
\nn R_2 &=& -d(K_R dX_R)~,\\
\nn S_1 &=& d(K_T J d\chi + K_LJdX_L-K_RJdX_R)~,\\
\nn S_2 &=&- d(K_C J d\phi + K_LJdX_L + K_RJdX_R)~,
\ee
where $d$ is the de Rham differential and $J$ is the trivial diagonal
complex structure so that 
$K_L J d\phi$ is short-hand for $i K_l dX_l - i K_{\bar{l}}d \bar X_{\bar l}$ and so on.
Here $f$ and $g$ depend only on $\boldsymbol{\phi}$ and 
$\boldsymbol{\chi}$ respectively and are introduced to take care of the
ambiguity in the definition of $N_{1,2}$
Namely, under a change of coordinates $\boldsymbol{\phi}^\prime 
(\boldsymbol{\phi})$ and $\boldsymbol{\chi}^\prime(\boldsymbol{\chi})$ the 
exponentials
$e^{f(\boldsymbol{\phi})}$ and $e^{g(\boldsymbol{\chi})}$
transform as densities,
\be
\label{transformfandg}
  e^{f^\prime(\boldsymbol{\phi}^\prime)} = e^{f (\boldsymbol{\phi})} \det \left ( \frac
  {\partial \boldsymbol{\phi}^\prime}{\partial
   \boldsymbol{\phi}}\right )~,~~~~~~~~
  e^{g^\prime(\boldsymbol{\chi}^\prime)} = e^{g (\boldsymbol{\chi})} \det \left ( \frac
  {\partial \boldsymbol{\chi}^\prime}{\partial
  \boldsymbol{\chi}}\right )~.
\ee
so that $N_{1,2}$ themselves are invariant.

First of all one has to check that these are the pure spinors which annihilate the correct 
subspace of $(T\oplus T^*)\times \mathbb{C}$, i.e. those subspace defined by the 
generalized 
complex structures ${\cal J}_1$ and ${\cal J}_2$. The proof of this statement involves
the Gualtieri map and is technical. We give the details in appendix
\ref{appendix2}.
The pure spinors are trivially closed $d\rho_{1,2} = 0$ and moreover
$(\rho_1,\rho_2)=(\rho_1,\bar{\rho}_2) = 0$ (see appendix \ref{appendix2}).
The only remaining 
condition for generalized Calabi-Yau metric is  $(\rho_1,\bar{\rho}_1) = \alpha 
(\rho_2,\bar{\rho}_2)$ for some positive constant $\alpha$. Explicitly, this condition
becomes
\be
\label{bigPfaffianMA}
(-1)^{\frac{d_c(d_c-1)}{2}} e^{f(\boldsymbol{\phi})} e^{\bar{f}(\bar{\boldsymbol{\phi}})}~
{\rm Pf}\left(\begin{array}{cccccc}
 0 & - K_{l\bar{l}} & - K_{lr} & 0 & 0 & - K_{l\bar{t}} \\
 K_{\bar{l}l} & 0 & 0 &  K_{\bar{l}\bar{r}}  &  K_{\bar{l}t} & 0 \\
 K_{rl} & 0 & 0 &  K_{r\bar{r}} &  K_{rt} & 0 \\
 0 & - K_{\bar{r}\bar{l}} & - K_{\bar{r}r} & 0 & 0 & -K_{\bar{r}\bar{t}} \\
  0 & - K_{t\bar{l}} & - K_{tr} & 0 & 0 & - K_{t\bar{t}} \\
    K_{\bar{t}l} & 0 & 0 &  K_{\bar{t}\bar{r}} &  K_{\bar{t}t} & 0 
\end{array}\right)\\
=(-1)^{\frac{d_t(d_t-1)}{2}}\alpha~   e^{g(\boldsymbol{\chi})} e^{\bar{g}
(\bar{\boldsymbol{\chi}})}~  
   {\rm Pf}\left(\begin{array}{cccccc}
  0 &  K_{l\bar{l}} & 0  & K_{l\bar{r}} & 0 &  K_{l\bar{c}} \\
- K_{\bar{l}l} &  0 &  - K_{\bar{l}r} & 0 & - K_{\bar{l}c} & 0 \\
  0 &  K_{r\bar{l}} & 0 &  K_{r\bar{r}}  & 0 &  K_{r\bar{c}}\\
 - K_{\bar{r}l} & 0 & -K_{\bar{r}r} & 0 & -K_{\bar{r}c} & 0 \\
  0 &  K_{c\bar{l}} & 0 &  K_{c\bar{r}} & 0 &  K_{c\bar{c}}\\
 - K_{\bar{c}l} & 0 &- K_{\bar{c}r} & 0 & - K_{\bar{c}c} & 0  
\end{array}\right)\nn
\ee
where we used the definition (\ref{mukaipairing}) of the Mukai pairing and
the usual definition the Pfaffian (see (\ref{pfaff}) below). 

For a $2n \times 2n$ antisymmetric matrix
${\cal A}=\{a_{\mu\nu}\}$ one associates a two-form 
\be
  a = \sum\limits_{\mu < \nu} a_{\mu\nu} ~e^\mu \wedge e^\nu
\ee
with the standard basis $\{ e^1, e^2, ... , e^{2n}\}$ in $\mathbb{R}^n$. The Pfaffian is 
defined as
\be\label{pfaff}
 \frac{1}{n!} a^n  = {\rm Pf} ({\cal A})~ e^1 \wedge e^2 \wedge ... \wedge e^{2n}~,
\ee
where $a^n$ denotes the wedge product of $n$ copies of $a$ with itself.

Our conventions are such that $\alpha$ in (\ref{twospinorsCYrel}) is positive, assuming a 
Riemannian metric $g_{\mu\nu}$. If $d_c \neq d_t$ 
then the constant $\alpha$ can be easily absorbed by a rescaling of $K$, thus giving 
the same geometry (a rescaling of the generalized \kah potential will just rescale all
geometric objects). However, in the case $d_c = d_t$ (including the
case $d_c = d_t =0$) this is not possible. Instead, if $d_c = d_t > 0$ we may rescale
the $\phi$ or $\chi$ coordinates to get $\alpha = 1$. Only in the case $d_c=d_t = 0$
this is not possible and we get a family of PDEs parametrized by alpha.
  
There are two useful relations for the
Pfaffian we can use to simplify further the condition (\ref{bigPfaffianMA}). 
For $2n \times 2n$ antisymmetric matrix ${\cal A}$ and an arbitrary $2n \times 2n$ 
matrix $V$ we have the relation
\be
  \label{PfdetPfrelation}
     {\rm Pf} (V{\cal A}V^T) = \det (V) ~{\rm Pf}({\cal A})~. 
\ee
For an arbitrary $n\times n$-matrix $M$ we may write
 \be
 \label{Pfafproperty}
  {\rm Pf} \left ( \begin{array}{cc}
   0 & M \\
   -M^T & 0 \end{array} \right ) = (-1)^{n(n-1)/2} \det M~.
 \ee
Using the relation (\ref{PfdetPfrelation}) we can reorder rows and columns by
choosing an appropriate $V$ and arrive at a block off-diagonal form. To this we
can apply (\ref{Pfafproperty}) arriving at
\be
 \label{generalMAdet}
 (-1)^{d_s d_c}  e^{f(\boldsymbol{\phi})} e^{\bar{f}(\bar{\boldsymbol{\phi}})} \det 
 \left ( \begin{array}{ccc}
  - K_{l\bar{l}} & - K_{lr} & - K_{l\bar{t}} \\
  - K_{\bar{r}\bar{l}} & - K_{\bar{r}r} & - K_{\bar{r}\bar{t}} \\
  - K_{t\bar{l}} & -K_{tr} & -K_{t\bar{t}} \end{array} \right ) = \alpha 
  e^{g(\boldsymbol{\chi})} e^{\bar{g}(\bar{\boldsymbol{\chi}})}  \det \left (\begin
  {array}{ccc}
   K_{l\bar{r}} & K_{l\bar{l}} & K_{l\bar{c}} \\
   K_{r\bar{r}} & K_{r\bar{l}} & K_{r\bar{c}} \\
    K_{c\bar{r}} & K_{c\bar{l}} & K_{c\bar{c}}
    \end{array} \right )~,
\ee
where we have redefined the proportionality constant $\alpha$ (which is still positive)
and chosen the minuses such that the determinants reduce to determinants of
positive definite matrices in known cases.
We refer to this relation as the generalized Monge-Amp\`ere equation. If the 
generalized \kah potential $K$ satisfies this non-linear PDE then $K$ gives rise to a 
generalized Calabi-Yau metric structure. For the non-degeneracy of the metric it is 
important  that the determinants in (\ref{generalMAdet}) are nowhere zero. 

Using the relation (\ref{twospinorsCYrel}) we can write down an explicit formula for 
the dilaton $\Phi$. 
We need to calculate the determinant of the metric in our coordinates.   
Using the the identities from appendix \ref{appendix1} it is lengthy but
straightforward to calculate the determinant of the metric\footnote{The calculation
of the determinant of the metric results in an absolute square. When taking the square root of the determinant the sign has been
determined so that the square root is positive when the generalized Monga-Amp\`ere
equation is satisfied.}
\be
\label{detg}
\sqrt{\det g_{\mu\nu}} = \frac{(-1)^{d_sd_c}}{\det K_{LR}}
  \det \left ( \begin{array}{ccc}
  - K_{l\bar{l}} & - K_{lr} & - K_{l\bar{t}} \\
  - K_{\bar{r}\bar{l}} & - K_{\bar{r}r} & - K_{\bar{r}\bar{t}} \\
  - K_{t\bar{l}} & -K_{tr} & -K_{t\bar{t}} \end{array} \right ) 
   \det \left (\begin
  {array}{ccc}
   K_{l\bar{r}} & K_{l\bar{l}} & K_{l\bar{c}} \\
   K_{r\bar{r}} & K_{r\bar{l}} & K_{r\bar{c}} \\
    K_{c\bar{r}} & K_{c\bar{l}} & K_{c\bar{c}}
    \end{array} \right )~.
\ee
It should be stressed that the determinant of the metric is not invariant under 
coordinate transformations; it of course transforms as a density. It is therefore
important to emphasize that the formula above is given in coordinates defined by the
superfields and that in the case with semichiral superfields these coordinates are
not adapted to either of the complex structures. 
Using (\ref{twospinorsCYrel})  and (\ref{detg}) we get an expression for the
dilaton $\Phi$
\be
\label{dilaton}
e^{2\Phi} &=& (-1)^{d_s d_c}\frac{ e^{-f(\boldsymbol{\phi})} 
e^{-\bar{f}(\bar{\boldsymbol{\phi}})}}{\det K_{LR}}
  \det \left ( \begin{array}{ccc}
  - K_{l\bar{l}} & - K_{lr} & - K_{l\bar{t}} \\
  - K_{\bar{r}\bar{l}} & - K_{\bar{r}r} & - K_{\bar{r}\bar{t}} \\
  - K_{t\bar{l}} & -K_{tr} & -K_{t\bar{t}} \end{array} \right ) \\
\nn &=&
\frac{ e^{-g(\boldsymbol{\chi})} e^{-\bar{g}(\bar{\boldsymbol{\chi}})}}{\alpha \det K_{LR}}
   \det \left (\begin
  {array}{ccc}
   K_{l\bar{r}} & K_{l\bar{l}} & K_{l\bar{c}} \\
   K_{r\bar{r}} & K_{r\bar{l}} & K_{r\bar{c}} \\
    K_{c\bar{r}} & K_{c\bar{l}} & K_{c\bar{c}}
    \end{array} \right )~.
\ee

An important issue is how these relations behave under coordinate changes. For
instance, under diffeomorphisms that preserve our superspace structure
(so that coordinates associated with a particular superfield mixes only with coordinates
associated to superfields that satisfy the same superspace constraints), i.e. diffeomorphisms satisfying $\boldsymbol{X}_l^\prime
(\boldsymbol{X}_l, \boldsymbol{\phi},
\boldsymbol{\chi})$, 
$\boldsymbol{X}_r^\prime(\boldsymbol{X}_r, \boldsymbol{\phi},\bar{\boldsymbol{\chi}})
$, $\boldsymbol{\phi}^\prime(\boldsymbol{\phi})$,
and $\boldsymbol{\chi}^\prime(\boldsymbol{\chi})$, it is possible
to show that
\be
  \det \left ( \begin{array}{ccc}
  - K_{l\bar{l}} & - K_{lr} & - K_{l\bar{t}} \\
  - K_{\bar{r}\bar{l}} & - K_{\bar{r}r} & - K_{\bar{r}\bar{t}} \\
  - K_{t\bar{l}} & -K_{tr} & -K_{t\bar{t}} \end{array} \right ) 
\ee
transforms with a factor
\be
\det \left ( \frac{\partial \boldsymbol{X}_l^\prime}{\partial \boldsymbol{X}_l} \right )
\det \left ( \frac{\partial \boldsymbol{X}_r^\prime}{\partial \boldsymbol{X}_r} \right )
\det \left (\frac{\partial \boldsymbol{\chi}^\prime}{\partial \boldsymbol{\chi}} \right ) 
\times \;c.c.
\ee
while
\be
  \det \left (\begin
  {array}{ccc}
   K_{l\bar{r}} & K_{l\bar{l}} & K_{l\bar{c}} \\
   K_{r\bar{r}} & K_{r\bar{l}} & K_{r\bar{c}} \\
    K_{c\bar{r}} & K_{c\bar{l}} & K_{c\bar{c}}
    \end{array} \right )
\ee
transforms with a factor
\be\label{transA}
\det  \left ( \frac{\partial \boldsymbol{X}_l^\prime}{\partial \boldsymbol{X}_l} \right ) 
\det \left ( \frac{\partial \boldsymbol{X}_r^\prime}{\partial \boldsymbol{X}_r} \right )
\det \left ( \frac{\partial \boldsymbol{\phi}^\prime}{\partial \boldsymbol{\phi}} \right )
\times \;c.c.
\ee
In particular this means that  the generalized 
Monge-Amp\`ere equation (\ref{generalMAdet}) is invariant under these changes
of coordinates if we take into account
how $e^{f(\boldsymbol{\phi})}$ and $e^{g(\boldsymbol{\chi})}$ transform, see 
(\ref{transformfandg}).
We can also investigate how the dilaton changes under coordinate transformations.
Since $\det K_{LR}$ transforms with a factor
\be
\det \left ( \frac{\partial \boldsymbol{X}_l^\prime}{\partial \boldsymbol{X}_l} \right )
\det  \left ( \frac{\partial \boldsymbol{X}_r^\prime}{\partial \boldsymbol{X}_r} \right ) 
\times\; c.c.
\ee 
and using the transformation of (\ref{transA}) we see that the dilaton does not change. 
Again we should use the correct transformations (\ref{transformfandg})
for $e^{f(\boldsymbol{\phi})}$ and $e^{g(\boldsymbol{\chi})}$. Indeed this is what
we expect since the dilaton $\Phi$ is a function.

It follows from (\ref{transformfandg}) that we can always change to a coordinate system 
such that $f=g=1$ in a given coordinate patch. We do this in what follows, simplifying the 
generalized Monge-Amp\`ere equation (\ref{generalMAdet}). Furthermore, for the case 
$d_c+d_t\neq 0$ we can always choose the constant $\alpha$ to be 1. However, in the 
case $d_c=d_t=0$ we cannot remove the constant $\alpha$. So in the generic case
where $d_c+d_t\neq 0$ the generalized Monge-Amp\`ere equation becomes
\be
\label{simplegMA}
 (-1)^{d_s d_c} \det 
 \left ( \begin{array}{ccc}
  - K_{l\bar{l}} & - K_{lr} & - K_{l\bar{t}} \\
  - K_{\bar{r}\bar{l}} & - K_{\bar{r}r} & - K_{\bar{r}\bar{t}} \\
  - K_{t\bar{l}} & -K_{tr} & -K_{t\bar{t}} \end{array} \right ) = 
 \det \left (\begin
  {array}{ccc}
   K_{l\bar{r}} & K_{l\bar{l}} & K_{l\bar{c}} \\
   K_{r\bar{r}} & K_{r\bar{l}} & K_{r\bar{c}} \\
    K_{c\bar{r}} & K_{c\bar{l}} & K_{c\bar{c}}
    \end{array} \right )~,
\ee
and the dilaton becomes
\be
\label{simpleDilaton}
e^{2\Phi} &=& (-1)^{d_s d_c}\frac{1}{\det K_{LR}}
  \det \left ( \begin{array}{ccc}
  - K_{l\bar{l}} & - K_{lr} & - K_{l\bar{t}} \\
  - K_{\bar{r}\bar{l}} & - K_{\bar{r}r} & - K_{\bar{r}\bar{t}} \\
  - K_{t\bar{l}} & -K_{tr} & -K_{t\bar{t}} \end{array} \right ) \\
\nn &=&
\frac{1}{\det K_{LR}}
   \det \left (\begin
  {array}{ccc}
   K_{l\bar{r}} & K_{l\bar{l}} & K_{l\bar{c}} \\
   K_{r\bar{r}} & K_{r\bar{l}} & K_{r\bar{c}} \\
    K_{c\bar{r}} & K_{c\bar{l}} & K_{c\bar{c}}
    \end{array} \right )~.
\ee
which implies that for this choice of coordinates
\be
\label{simpleDilg}
e^{-4\Phi}\sqrt{\det g}=\det K_{LR}~.
\ee

Apart from the issue of the covariance of (\ref{generalMAdet}), there are other
ambiguities in finding the pure spinors. First of all,  the relation
between the  form of the pure spinors $\rho_1$, $\rho_2$ and the bihermitian
data is not one to one; there are ambiguities both in the definition of the pure
spinors and in the Gualtieri map. In particular, an overall $B$-transform gives
different but equivalent pure spinors. Also, our expression for the pure spinors
uses only closed forms so the pure spinors themselves are trivially closed. The
nontrivial condition comes from the normalization of the spinors. One could
alternatively choose to include a normalization factor in the definition of the
pure spinors so that they would be automatically normalized to $1$. Then
they would however not be automatically closed and the nontrivial condition
would come from imposing closedness.
When all is taken into account, these ambiguities do not effect the 
final answer (\ref{generalMAdet}).

\section{Special cases}
\label{special-cases}

In this section we consider a few special cases of the general equation
(\ref{generalMAdet}). We will see that some  special cases of
(\ref{generalMAdet}) have already appeared in the literature. 

\subsection{K\"ahler case: $d_s=d_t=0$} 

If $J_+ = J_-$ then $H=0$ and the generalized \kah manifold is a \kah manifold.
In this case $d_t=d_s=0$ and the pure spinors (\ref{definitionpurespinor})  are
\be
\rho_1 &=& d\boldsymbol{\phi}^1\wedge \ldots\wedge d\boldsymbol{\phi}^{d_c}~,\\
\rho_2 &=& e^{-id(K_C J d\phi)}~.
\ee
The equation (\ref{generalMAdet}) becomes 
\be
\label{393kkd03}
   \det (K_{c\bar{c}}) =1~,
\ee 
where we have absorbed $\alpha$ in a redefinition of $K$. This is the well-known
complex Monge-Amp\`ere equation implying Ricci flatness of the K\"ahler manifold.
In this case the dilaton $\Phi$ is constant. 

\subsection{Symplectic case: $d_c = d_t =0$}

In the case when $d_c=d_t=0$ both generalized complex structures ${\cal J}_1$ and
${\cal J}_2$ are of symplectic type. In this case the pure spinors become
\be
\rho_1 &=& e^{-d(K_LdX_L)+ id(K_LJdX_L-K_RJdX_R)}~,\\
\rho_2 &=& e^{-d(K_RdX_R)-id(K_LJdX_L+K_RJdX_R)}~.
\ee
The  generalized Monge-Amp\`ere equation (\ref{generalMAdet}) becomes 
\be
 \label{semiMAdet}
  \det \left ( \begin{array}{cc}
  K_{l\bar{l}} &  K_{lr}  \\
  K_{\bar{r}\bar{l}} &  K_{\bar{r}r} 
   \end{array} \right ) = \alpha~ \det \left (\begin{array}{cc}
   K_{l\bar{r}} & K_{l\bar{l}}  \\
   K_{r\bar{r}} & K_{r\bar{l}} 
    \end{array} \right )~.
\ee
Here $\alpha$ cannot be removed neither by  rescaling of $K$ nor by a rescaling of  
coordinates and thus we are dealing with a family of equations. 

To compute the dilaton in this case is straightforward. Using (\ref{dilaton}) we get the 
following expression for 
the dilaton $\Phi$
\be
 \label{dilatononlysemis}
  \Phi = \frac{1}{2} \ln  \frac{\det \left ( \begin{array}{cc}
  K_{l\bar{l}} &  K_{lr}  \\
  K_{\bar{r}\bar{l}} &  K_{\bar{r}r} 
   \end{array} \right )}{\det \left ( \begin{array}{cc}
  K_{l r} &  K_{l\bar{r}}  \\
  K_{\bar{l} r} &  K_{\bar{l}\bar{r}} 
   \end{array} \right )}~.
\ee

For the special case $d_s = 1$ (\ref{semiMAdet}) collapses to 
\be
 K_{rl} K_{\bar{r}\bar{l}} + \alpha K_{r \bar{l}} K_{\bar{r}l} = 
 (1+ \alpha) K_{l\bar{l}} K_{r\bar{r}}~,
\ee
which is equivalent to the statement that 
\be
 \label{anticomcomplexst}
 J_+ J_- + J_- J_+ =2 \left ( \frac{1-\alpha}{1+\alpha} \right  ) \mathbb{I} 
\ee
and, as shown in \cite{Lindstrom:2005zr}, when $ \left ( \frac{1-\alpha}{1+\alpha} 
\right  )^2<1$ (which is true since $\alpha$ is positive), it is possible
to explicitly construct a hyper\kah structure so the geometry is hyper\kah.
See  \cite{Goteman:2009xb} 
for further discussion.  For general $d_s$, (\ref{anticomcomplexst}) becomes
\be
  J_+J_-+J_-J_+ = 2
  \left(\frac{1-\alpha^{\frac{1}{d_s}}}{1+\alpha^{\frac{1}{d_s}}}\right)
    \mathbb{I}
\ee
which similarly implies that
the geometry is hyperK\"ahler and it automatically solves (\ref{semiMAdet}).
We have checked that the expression (\ref{dilatononlysemis})  for the dilaton reduces to a
constant when the geometry is hyper\kah.

\subsection{$[J_+, J_-]=0$: $d_s=0$}

Consider the case when $d_s=0$ and thus $[J_+, J_-]=0$. This manifold admits a local 
product structure. The pure spinors become
\be
\rho_1 &=& d\boldsymbol{\phi}^1\wedge \ldots\wedge d\boldsymbol{\phi}^{d_c}
~  e^{id(K_T Jd\chi)}~,\\
\rho_2 &=& d\boldsymbol{\chi}^1\wedge \ldots\wedge d\boldsymbol{\chi}^{d_t}
~  e^{-id(K_ C Jd\phi)}~.
\ee
The generalized  Monge-Amp\`ere equation (\ref{generalMAdet}) collapses to 
\be
\label{genMAct}
\det (K_{c\bar{c}}) =  \det(- K_{t\bar{t}})~,
\ee
where $\alpha$ can be absorbed either in rescaling of $K$ or in the rescaling of $\chi$-
direction. In this case we have  
\be
  \sqrt{g} = \det (K_{c\bar{c}}) \det (- K_{t\bar{t}})
\ee
and thus from (\ref{twospinorsCYrel}) the dilaton is
\be
   \Phi = \frac{1}{2} \ln \det (K_{c\bar{c}}) ~.
\ee
This equation first appeared in \cite{Buscher:1985kb}
(see also \cite{Hull:1986iu} and \cite{Rocek:1991ze}). 

For the case $d_c=d_t=1$  we get a linear equation which is equivalent to the
requirement of $N=(4,4)$ supersymmetry for the corresponding sigma model. Indeed
the general conditions for $N=(4,4)$ supersymmetry for models with $d_s=0$
and $d_c = d_t$ \cite{Gates:1984nk} can be written in matrix form as follows 
\be
\label{N=4cta} &&  K_{c\bar{c}} + K_{t\bar{t}} = 0 \\
\label{N=4ctb}  &&  ( K_{c\bar{c}})^T = \overline{K_{c\bar{c}}}
\ee
which certainly gives a solution of (\ref{genMAct}).
In fact   we  need only condition (\ref{N=4cta}) to solve the generalized
Monge-Amp\`ere equation (\ref{genMAct}). Thus we can produce many solutions of
this non-linear PDE. 

\subsection{$d_t=0$}

Let us finally comment on the case when $d_t=0$ and thus one of the generalized 
complex structure is of symplectic type. The corresponding pure spinors are 
\be
\rho_1 &=& d\boldsymbol{\phi}^1\wedge \ldots\wedge d\boldsymbol{\phi}^{d_c}
~e^{-d(K_LdX_L)+id(K_LJdX_L-K_RJdX_R)}~,\\
\rho_2 &=& e^{-d(K_RdX_R)-id(K_C Jd\phi + K_LJdX_L+K_RJdX_R)}~.
\ee
This case has been studied previously by Halmagyi and Tomasiello in
\cite{Halmagyi:2007ft}, especially for $D=6$ due to its relevance for supergravity.
Their pure spinors differ by a $b$-transform from the solution presented
here. However their final generalized Monge-Amp\'ere equation is the same as ours
for this special case.

The explicit form of the dilaton $\Phi$ is provided by (\ref{dilaton}). 
 
\section{Geometrical considerations}
\label{geometry}

In this section we want to elaborate on a few geometrical points related to the
generalized Monge-Amp\'ere equation
(\ref{generalMAdet}). We are not going to derive any new results, however it is
instructive to consider the classical differential geometry   of
our equations. Moreover it provides
a strong consistency check of our formulas. 
  
The existence of a generalized Calabi-Yau metric structure is equivalent to
the existence of spinors $\epsilon^\pm$ on the generalized Calabi-Yau space satisfying
the
following supersymmetry equations
\be
\label{susy1}
\nabla^{(\pm)} \epsilon^{(\pm)} &=& 0~,\\
\label{susy2} ( d\Phi \pm \frac{1}{2} H ) \epsilon^{(\pm)} &=& 0~.
\ee 
The equation (\ref{susy1})  is the so called gravitino equation while equation 
(\ref{susy2}) is the dilatino equation. 
Here $\nabla^{(\pm)}$ are the covariant derivatives with connections with torsion 
$\Gamma^{(\pm)} = \Gamma \pm \frac 1 2 g^{-1} H$ where $\Gamma$ is the usual 
Christoffel symbol, while (\ref{susy2}) involves the Clifford action of the one-form $d\Phi$ 
and the 3-form $H$ on the spinors. See \cite{Witt:2006tq} for further details of the 
notation and the relation of these equations to generalized Calabi-Yau structures.
  
On a generalized \kah manifold the equation (\ref{susy1}) implies that the holonomies
of the connections $\nabla^{(\pm)}$ are  both in $SU(n)$  and the equation (\ref{susy2})
implies a relation between the dilaton $\Phi$ and the rest of the geometrical
data $(g, H, J_\pm)$, which is discussed below. Now we want to 
impose these conditions on the generalized \kah geometry. 
   
In the following we use formulas from \cite{Hull:1986iu},\cite{Hull:1985zy}
and \cite{Hull:XXX}. 
We define the $U(1)$ parts of the
connections $\Gamma^{(\pm)}_\mu$ 
\be
\Gamma^{(\pm)}_\rho = J^\mu_{\pm \nu}\Gamma^{(\pm)\nu}_{\rho\mu} \, ,
\ee
the $U(1)$ parts of the curvatures
\be
C^{(\pm)}_{\mu\nu} = J_{\pm\lambda}^\rho R^\lambda_{\;\rho\mu \nu}
\ee
and the $J$-trace of $H$
\be
v^{(\pm)}_\rho  = \pm
  J_{\pm \rho}^{\mu}  H_{\mu \nu \sigma} J_{\pm}^{\nu\sigma}
=2J_{\pm \rho}^{\mu}\nabla_\nu J^\nu_{\pm \mu} 
~.
\ee
For formulae involving containing $v^{(+)}$
or $\Gamma^{(+)}$
it is useful to  use complex coordinates adapted to $J_+$
while for formulae involving $v^{(-)}$ or $\Gamma^{(-)}$ we use 
complex coordinates adapted to $J_-$.
In both cases, we use the indices $a,b...$ to label holomorphic coordinates and 
indices $\bar{a},\bar{b},...$ to  to label anti-holomorphic coordinates.
Then in such complex coordinate systems, we have
\be
\Gamma^{(\pm)}_a  = 
i\left(\Gamma_{ac}^{(\pm)c}
-\Gamma_{a\bar c}^{(\pm)\bar c}\right) 
\, ,  \qquad
v^{(\pm)}_a  =
-\Gamma_{a\bar c}^{(\pm)\bar c} 
= \mp 2H_{a\bar c}^{(\pm)\bar c}~,
\ee
where $H^{(\pm)}_{ab\bar c}$ is the $(2,1)$ part of the three form $H$ with respect
to $J_\pm$, 
and
\be
C^{(\pm)}_{\mu\nu}  =
\partial_{\mu} \Gamma^{(\pm)}_{\nu}-
\partial_{\nu} \Gamma^{(\pm)}_{\mu}~.
\ee
The $U(1)$ connection can be expressed in terms of the determinant of the metric
and the one form $v^{(\pm)}$ as \cite{Hull:1985zy}
\be
\label{sflskdjf}
\Gamma^{(\pm)}_a = i\left(2v^{(\pm)}_a + \partial_a\ln \sqrt{\det g_{\mu\nu}}\right)
\ee

Equation (\ref{susy1}) is equivalent to the holonomy of the $\nabla^{(\pm)}$
connections both being in $SU(n)$,
i.e. $C^{(\pm)}_{\mu\nu}=0$.  At the same time, equation (\ref{susy2})
implies that $v^{(\pm)} = - 2 d\Phi$ and thus $\Gamma^{\pm}$ can be written
as
\be
  \label{u1conn}
\Gamma^{(\pm)}_a = i\partial_a\ln \left( e^{-4\Phi}\sqrt{\det g}\right)~,
\ee
using the complex coordinates adapted to $J_+$ for $\Gamma^{(+)}$ and those
adapted to $J_-$ for $\Gamma^{(-)}$.
Thus using these formulas we arrive at the statement that 
the equations (\ref{susy1}) and (\ref{susy2}) are equivalent the following conditions \cite{Hull:1986iu}
\be
   \label{MAsugrahhq}
&&   \partial \bar{\partial}\ln \left( e^{-4\Phi}\sqrt{\det g}\right ) = 0~,\\
\label{38383} &&  \label{893usus}  v^\pm  = - 2d\Phi ~,
\ee
where there are two equations of the form (\ref{MAsugrahhq}), one with $J_+$
complex coordinates and one with $J_-$ complex coordinates. Then equation
(\ref{MAsugrahhq}) implies that locally we can choose $J_+$ complex or $J_-$
complex coordinates such that
\be
\label{3u3e9933}
e^{-4\Phi}\sqrt{\det g} = 1\;.
\ee
The transformations relating the complex coordinates adapted to $J_+$ or $J_-$ to
our superspace-inspired coordinates $(\phi,\chi,X_L,X_R)$ are given in Appendix
\ref{appendix1}. Using these, the equations given by (\ref{3u3e9933}) in $J_+$ 
coordinates and by (\ref{3u3e9933}) in $J_-$ coordinates both give the same equation
in $\phi,\chi,X_L,X_R$ coordinates, and that is
\be
e^{-4\Phi}\sqrt{\det g} = \det K_{LR}~.
\ee
and we have seen in (\ref{simpleDilg}) that there always exist coordinates
where the above equation is identically satisfied\footnote{This provides a
proof that the generalized Monge-Amp\`ere
equation implies (\ref{susy1}) and (\ref{susy2}) or equivalently 
(\ref{MAsugrahhq}) and (\ref{38383}).
We believe that   the converse is also true, namely that the
equations
(\ref{MAsugrahhq}) and (\ref{38383}) imply our generalized Monge-Amp\`ere equation.
At present we have only been able to prove this for the case where $d_s=0$
in which case there is a simple formula for $v^{(\pm)}$ in terms of the generalized
\kah potential.}.

Within the present framework it is easy to show that we are dealing with the solution of 
the equation (\ref{oneloopeqsmotion}).  
Let us work in $J_+$ complex coordinates. In this case the Ricci
tensor can be written as follows \cite{Hull:1985zy}
\be
\label{eqcomep22}
R_{ab}^{(+)} &=& \nabla_a^{(-)} v^{(+)}_b~,
\\
\label{eqcomep33} R_{a\bar b}^{(+)} &=& \nabla_a^{(-)}v^{(+)}_{\bar b}
-i C^{(+)}_{a\bar b}-2 (\partial _a v^{(+)}_{\bar b}- \partial _{\bar b} v^{(+)}_a)
~.
\ee
Using equation (\ref{3u3e9933}) we find that  the $U(1)$
connection $\Gamma^{(+)}_\mu$ is zero, so $C^{(+)}_{a\bar b}=0$.
Thus substituting $v^{(+)} = - 2d\Phi$ into (\ref{eqcomep22}) and (\ref{eqcomep33}) 
we arrive at equation (\ref{oneloopeqsmotion}). 
Then the conditions for supersymmetry
imply the equations of motion are satisfied, as was to be expected.

\section{Physical interpretation} 
\label{physics}

 Generalized \kah geometry is the target space geometry of
classical $N=(2,2)$ supersymmetric sigma models. 
In this section we  comment on the physical interpretation of our generalized
Monge-Amp\`ere equation  in relation to the finiteness, conformal invariance and supersymmetry of such 
sigma models.

A necessary condition for such a sigma model to be conformally invariant at one-loop
is\footnote{This ensures that the dilaton beta-function is a constant at one loop. Full conformal invariance requires that this constant contribution is cancelled by contributions from ghosts and other matter fields.}
\be
\label{1loopbeta}
 R^{(+)}_{\mu\nu} + 2 \nabla^{(-)}_\mu \partial_\nu \Phi = 0~.
\ee
The condition for one-loop finiteness is however the weaker condition that \cite{Hull:1985rc}
\be
\label{1loopfin}
 R^{(+)}_{\mu\nu} +  \nabla^{(-)}_\mu V_\nu + 2\partial _{[\mu} W _{\nu]}= 0
\ee
for some $V_\mu, W_\mu$. This is sufficient to ensure that the one-loop counterterm is a total derivative when the classical equations of motion are used.
Finally, we have seen that such a sigma model has target space supersymmetry if
\be\label{sumetric1}
C^{(\pm)}_{\mu\nu}=  0~,\\
\label{sumetric2}
v^{(\pm)} = -2d\Phi~.
\ee

We now consider the implications of these for our geometries, using (\ref{eqcomep22}),(\ref{eqcomep33})
and
\be
\label{chol}
C^{(\pm)}_{ab}=4i \partial _{[a}v^{(\pm)}_{b]} \,  ,
\ee
which follows from (\ref{sflskdjf}).
Consider first the condition (\ref{sumetric1}) for $SU(n)$ holonomy, $C^{(\pm)}_{\mu\nu}=  0$.
The condition (\ref{1loopfin}) for one-loop finiteness was analysed in this case in \cite{Hull:1986iu}
 and is satisfied with
 $W_\mu=2 v^{(+)}_\mu$ and $V_\mu=-v^{(+)}_\mu$.
From (\ref{chol}), $\partial _{[a}v^{(\pm)}_{b]}=0$, so that locally
\be
v^{(\pm)}_a=-2\partial _a (A^{(\pm)}+iB^{(\pm)})
\ee
for some $A^{(\pm)},B^{(\pm)}$.
Then using  (\ref{sflskdjf}) we have the Monge-Ampere equations \cite{Hull:1986iu}
\be
\partial\bar\partial\ln\left(e^{-4(A^{(\pm)}+iB^{(\pm)})}\sqrt{\det g}\right) = 0~.
\ee
For supersymmetry we need to impose in addition (\ref{sumetric2}), which implies
\be
A^{(\pm)}=\Phi,\qquad dB^{(\pm)}=0~.
\ee
We then recover our Monge-Ampere equations \cite{Hull:1986iu}
\be
\partial\bar\partial\ln\left(e^{-4\Phi}\sqrt{\det g}\right) = 0
\ee
and find that the condition (\ref{1loopbeta})
 for conformal invariance is satisfied. Thus the
 $SU(n)$ holonomy condition (\ref{sumetric1}) implies one-loop finiteness, and the second supersymmetry condition
 (\ref{sumetric2}) implies one-loop conformal invariance, so that supersymmetric backgrounds are conformal.

Next, we assume the conformal condition  (\ref{1loopbeta}).
Defining
\be
k_a ^{(\pm)}=v_a ^{(\pm)}+2\partial_a\Phi 
\ee
we find that
\be
\nabla^{(\mp)}_a k_b ^{(\pm)}= 0
\ee
and
\be
\label{eqcomep33}  
i C^{(\pm)}_{a\bar b}
&=& \nabla_a^{(\mp)}k^{(\pm)}_{\bar b}
-2 (\partial _a k^{(\pm)}_{\bar b}- \partial _{\bar b} k^{(\pm)}_a)~.
\ee
A Killing vector ${\cal K}^\mu$ preserves $H$ provided
${\cal K}^\mu H_{\mu\nu\rho}$ is a closed 2-form.
Then if $C^{(\pm)}_{a\bar b}=0$, we see that
$k ^{(\pm) \mu}$ are Killing vectors preserving $H$.
Further, if in addition  $C^{(\pm)}_{a b}=0$, the Killing vectors can be written locally in terms of potentials
\be
k^{(\pm)}_a=-2\partial _a (\Phi + A^{(\pm)}+iB^{(\pm)})~.
\ee

A nice 
example\footnote{Our discussion is local and we do not take into account global 
issues, in particular those related to no-go theorems in supergravity for
compactifications with fluxes.}
of this situation  would be $S^3 \times S^1$ which admits two different 
solutions of (\ref{1loopbeta}). For a given \kah potential \cite{Rocek:1991vk} we can 
identify two solutions for this geometry $(g_{\mu\nu}, H_{\mu\nu\rho}, 0)$
and $(g_{\mu\nu}, H_{\mu\nu\rho}, \Phi)$ with $\Phi$ given by (\ref{dilaton}).
The metric and $H$-field are the same for 
both solutions, but one has zero dilaton and the other one has non-zero dilaton. This 
unusual situation  occurs since the dilaton satisfies the following equation 
\be
  \nabla_\mu^{(-)} \partial_\nu \Phi = 0~
\ee
so that $K^\mu =g^{\mu\nu} \partial_\nu \Phi$ is a Killing vector preserving $H$.
  The solution
with non-zero dilaton
provides a supersymmetric supergravity solution (at least locally), while the one
with zero dilaton it will not.  However both backgrounds solve (\ref{1loopbeta}). 
A large class of non-comapct solutions with Killing vectors was given in \cite{Lust}.

As an alternative to the geometrical considerations, one can try to find the conditions for conformal invariance directly  by perforing a
one-loop calculation in $N=(2,2)$ 
superspace. However, in the presence of semi-chiral superfields, this is not so simple.
The main problem is again that there is no good understanding of the dilaton in the
general situation. Thus in \cite{Grisaru:1997pg} the authors only study
a necessary condition for conformal invariance, ultra-violet finiteness at one loop.
\footnote{
The value of the work  \cite{Grisaru:1997pg} is that it 
contains general expression for the counterterms of the $N=(2,2)$ sigma model,
while the precise relation between finiteness and conformality is not addressed,
see \cite{Hull:1985zy} for further comments regarding this issue. We hope to be
able to return to this question in a future publication}.
Comparing our
results with \cite{Grisaru:1997pg} 
we can see that the solutions of  the generalized Monge-Amp\' ere equation (\ref{generalMAdet}) are finite, as they should be.
However not every finite solution corresponds to the solution of (\ref{generalMAdet}). 
    
In order to show this we need to compare our result with \cite{Grisaru:1997pg}.
To this end we rewrite the generalized Monge-Amp\`ere equation using the 
well known property of block matrices
\be
 \det \left (\begin{array}{cc}
 A & B \\
 C & P \end{array}\right ) = \det (P) \det (A - B P^{-1} C )~,
\ee
where $P$ is assumed to be invertible. Then (\ref{generalMAdet}) can be rewritten 
as
\be
\label{UVfiniteness}
&&\frac{\det(K_{c\bar{c}})\det\left(\begin{array}{cc}
K_{l\bar{r}}-K_{l\bar{c}} K_{\bar{c}c}^{-1} K_{c\bar{r}} &
K_{l\bar{l}}- K_{l\bar{c}} K_{\bar{c}c}^{-1} K_{c\bar{l}}\\
K_{r\bar{r}}- K_{r\bar{c}} K_{\bar{c}c}^{-1} K_{c\bar{r}} &
K_{r\bar{l}} - K_{r\bar{c}} K_{\bar{c}c}^{-1} K_{c\bar{l}}\end{array}\right)}
{\det(-K_{t\bar{t}})\det\left(\begin{array}{cc}
-K_{l\bar{l}} + K_{l\bar{t}} K_{\bar{t}t}^{-1} K_{t\bar{l}} &
 -K_{lr} + K_{l\bar{t}} K_{\bar{t}t}^{-1}K_{tr}\\
-K_{\bar{r}\bar{l}} + K_{\bar{r}\bar{t}} K_{\bar{t}t}^{-1} K_{t\bar{l}} &
-K_{\bar{r}r}+ K_{\bar{r}\bar{t}} K_{\bar{t}t}^{-1} K_{tr}\end{array}\right)}  \\
\nn &&\\
\nn && = \alpha e^{-f(\boldsymbol{\phi})} e^{-\bar{f}(\bar{\boldsymbol{\phi}})}
   e^{g(\boldsymbol{\chi})} e^{\bar{g}(\bar{\boldsymbol{\chi}})}~,
\ee
where we have to assume the invertibility of $K_{c\bar{c}}$ and $K_{t\bar{t}}$.
Taking the logarithm  of the left hand side of (\ref{UVfiniteness}) we get exactly the
$N=(2,2)$ 
superspace counterterm calculated in \cite{Grisaru:1997pg}. Whenever the 
generalized Monge-Amp\`ere equation (\ref{generalMAdet}) is satisfied, the
counterterm vanishes  as the superspace integration  gives zero, which then implies one-loop
ultra-violet finiteness of the sigma model.
However, the general condition for one-loop finiteness is less restrictive than this: the  
counterterm
 vanishes  if the integrand 
 is given by  a 
generalized \kah gauge transformation, i.e.
\be
\label{UVfiniteness7474}
&& \frac{\det(K_{c\bar{c}})\det\left(\begin{array}{cc}
K_{l\bar{r}}-K_{l\bar{c}} K_{\bar{c}c}^{-1} K_{c\bar{r}} &
K_{l\bar{l}}- K_{l\bar{c}} K_{\bar{c}c}^{-1} K_{c\bar{l}}\\
K_{r\bar{r}}- K_{r\bar{c}} K_{\bar{c}c}^{-1} K_{c\bar{r}} &
K_{r\bar{l}} - K_{r\bar{c}} K_{\bar{c}c}^{-1} K_{c\bar{l}}\end{array}\right)}
{\det(-K_{t\bar{t}})\det\left(\begin{array}{cc}
-K_{l\bar{l}} + K_{l\bar{t}} K_{\bar{t}t}^{-1} K_{t\bar{l}} &
 -K_{lr} + K_{l\bar{t}} K_{\bar{t}t}^{-1}K_{tr}\\
-K_{\bar{r}\bar{l}} + K_{\bar{r}\bar{t}} K_{\bar{t}t}^{-1} K_{t\bar{l}} &
-K_{\bar{r}r}+ K_{\bar{r}\bar{t}} K_{\bar{t}t}^{-1} K_{tr}\end{array}\right)} \\
\nn &&\\
\nn && = F(\boldsymbol{\phi}, \boldsymbol{\chi}, \boldsymbol{X}_l)
   G(\boldsymbol{\phi}, \bar{\boldsymbol{\chi}}, \boldsymbol{X}_r)   F(\boldsymbol{\phi}, \boldsymbol{\chi}, \boldsymbol{X}_l)
   G(\boldsymbol{\phi}, \bar{\boldsymbol{\chi}}, \boldsymbol{X}_r)~.
\ee
Thus  we see again that  the generalized Monge-Amp\`ere equation
is more restrictive than just requiring one-loop ultra-violet finiteness. In section \ref{G-MA}
we discussed how these 
determinants transform. In the finiteness condition (\ref{UVfiniteness7474}) there is
no way to set $F$ and $G$ to one by an appropriate coordinate transformation
respecting the superfield structure.

\section{Summary}
\label{summary}
  
In this paper we have analyzed the local conditions for the existence of generalized 
Calabi-Yau metric structures. In the neighborhood of a regular point the geometry is 
characterized by a single function $K$ which is subject to a non-linear PDE, a 
generalization of the complex Monge-Amp\`ere equation. In deriving this equation 
we found an explicit local expression for the pair of closed pure spinors that encode the
geometry. 
   
One would clearly want to find solutions of the generalized Monge-Amp\`ere 
equation. Since we are dealing with a highly non-linear PDE, this task is very hard.
However, in some cases the generalized Monge-Amp\`ere equation can be 
replaced the by a collection of linear PDE's, which are stronger but easier to solve.
Namely, if the matrices on the left and right side of (\ref{generalMAdet}) are of the same 
size then we can require that they are related to each other by a similarity
transformation, e.g.,
by some constant matrix. This will give us a collection of linear PDE's and 
their solution will automatically satisfy the condition (\ref{generalMAdet}). Another 
approach would be to use sigma model dualities which would allow one to map the
non-linear PDE to a linear one, very much in analogy with
\cite{Hitchin:1986ea, Bogaerts:1999jc}. 

Another interesting physical aspect is to understand the proper relation between the 
conformal invariance of the $N=(2,2)$ supersymmetric sigma model and target space 
supersymmetry. In particular it would be important to understand the proper status of
the generalized 
Monge-Amp\`ere equation, since we know from the examples that it can accommodate 
not only supersymmetric supergravity solutions, but also non-supersymmertic
but conformal solutions. 
The presence of semi-chiral superfields complicate the situation and
we do not have anything to say about the matter at the moment. 

\bigskip\bigskip
\noindent{\bf\Large Acknowledgement}:
\bigskip\bigskip

\noindent

We thank  Alessandro Tomasiello, Frederik Witt  for discussions. 
We are grateful to the 2009
Simons Workshop where part of this work was carried out,
for providing a stimulating atmosphere.
The research of UL was supported by  VR grant 621-2009-4066.
The research of MR  was supported in part by NSF grant no.~PHY-06-53342.
The research of R.v.U. was supported by
the Simons Center for Geometry and Physics as well as the
Czech ministry of education under contract No.~MSM0021622409.
The research of M.Z. was supported by VR-grant  621-2008-4273.

\appendix

\Section{Appendix: Useful formulae from GK geometry}
\label{appendix1}
\setcounter{equation}{0}

In this appendix we review the relevant formulae for the local description of 
generalized \kah geometry in terms of the generalized \kah potential $K$. The
generalized \kah manifold is given by a bihermitian structure $(g, J_+, J_-)$ with the 
following integrability conditions
\be
   \label{generlkahintegrab}
      d^c_+ \omega_+ + d^c_- \omega_- =0~,~~~~~~~~d d^c_{\pm} \omega_\pm =0~, 
\ee
where $\omega_\pm = gJ_\pm$ and $d^c_\pm$ are $i(\bar{\partial}  -\partial)$   
operators associated  with the complex structures $J_\pm$. The NS-form $H$ is
defined from here as follows
\be
   H = d_+^c \omega_+ = - d^c_- \omega_-~,~~~~~~~~~dH=0~. 
\ee
In the neighborhood of a regular point (i.e., in neighbourhoods in which the 
rank of $(J_+  \pm J_-)$ is constant)
the  geometry can be solved in terms of a single real function $K$. Using the notation 
introduced in section \ref{G-MA} the complex structures $J_\pm$ are given 
by the following expressions
\ber
J_{+}=
\left(\begin{array}{cccc}
J_s &0&0&0\cr
K_{RL}^{-1} C_{LL}  &  K_{RL}^{-1}J_s K_{LR} &
K_{RL}^{-1}C_{LC}
& K_{RL}^{-1}C_{LT}\cr
0&0&J_c&0\cr
0&0&0&J_t\end{array}\right)~
\eer{Jplus}
and
\ber
J_{-}=\left(\begin{array}{cccc}
K_{LR}^{-1}J_s K_{RL} & K_{LR}^{-1}C_{RR} &
K_{LR}^{-1}C_{RC}& K_{LR}^{-1}A_{RT}\cr
0& J_s&0&0\cr
0&0&J_c&0\cr
0&0&0& -J_t\end{array}\right)~,
\eer{Jminus}
where $J_{s}$, $J_c$ and $J_t$ are canonical complex structures of size
$2d_s$, $2d_c$ and $2d_t$ respectively. 
Here we have also introduced the matrices $C$ and $A$ which are given by
the commutator or the anticommutator of the appropriate part of the hessian of $K$
with the canonical complex structure $J$:
$C_{\bullet\bullet} = J K_{\bullet\bullet} - K_{\bullet\bullet}J $, 
$A_{\bullet\bullet} = JK_{\bullet\bullet} +K_{\bullet\bullet}J$.
The expressions (\ref{Jplus}) and (\ref{Jminus}) are given in superspace inspired
coordinates $X_L,X_R,\phi,\chi$. In coordinates adapted to $J_+$, $J_+$ itself is
of course diagonal but $J_-$ is complicated. The form (\ref{Jplus}) we get by doing
a coordinate transformation from coordinates adapted to $J_+$ to the superspace
inspired coordinates. Similarly, in coordinates adapted to $J_-$, $J_-$ is diagonal and
the expression (\ref{Jminus}) is the coordinate transformation of that simple form
of $J_-$ from coordinates adapted to $J_-$ to superspace inspired coordinates,
see \cite{Lindstrom:2005zr}  for details. For our analysis we need the explicit
form of those coordinate transformation matrices.
If we call the coordinate basis vectors of the $J_+$ adapted coordinates $dX_L,dY_L,
d\phi,d\chi$, then the transformation matrix taking us from the superspace inspired
coordinate basis to the $J_+$ adapted coordinate basis is given by
\be
\left(\begin{array}{c}
dX_L\\ dY_L\\ d\phi\\ d\chi
\end{array}\right) = 
\left( \begin{array}{cccc}
1 & 0 & 0 & 0 \\
K_{LL} & K_{LR} & K_{Lc} & K_{Lt}\\
0 & 0 & 1&0\\
0 & 0 & 0 & 1
\end {array} \right)
\label{tardheosp3e90}
\left(\begin{array}{c}
dX_L\\ dX_R\\ d\phi\\ d\chi
\end{array}\right)~,
\ee
while if we call the coordinate basis vectors of the $J_-$ adapted coordinates
$dY_R,dX_R,d\phi,d\chi$, then the transformation matrix taking us from the superspace
inspired coordinate basis to the $J_-$ adapted coordinate basis is given by
\be
\left(\begin{array}{c}
dY_R\\ dX_R\\ d\phi\\ d\chi
\end{array}\right) = 
\left( \begin{array}{cccc}
K_{RL} & K_{RR} & K_{Rc} & K_{Rt}\\
0 & 1 & 0 & 0 \\
0 & 0 & 1&0\\
0 & 0 & 0 & 1
\end {array} \right)
\left(\begin{array}{c}
dX_L\\ dX_R\\ d\phi\\ d\chi
\end{array}\right)~.
\label{tardheosp3e91}
\ee
In this work however, the only thing that we will need is the determinant of those transformations. In particular the determinant of the metric will transform
as
\be
\label{deteh2922}  
\det g_{\mu\nu} ~~\longrightarrow~~   \left ( \det K_{LR} \right )^2~ \det g_{\mu\nu}~,
\ee
 when we pass from $J_\pm$-complex coordinates to our superfield coordinates. 

The integrability condition (\ref{generlkahintegrab}) can be solved locally in terms of 
two closed non degenerate two-forms \cite{Hull:2008vw}
\be
\label{relaGKtwosymprl}
{\cal F}_+ &=& \frac{1}{2}(B_+ - g)J_+ = \frac{1}{2} d \left ( -K_R J dX_R - K_C J
d\phi + K_T J d\chi \right )~,  \\
\label{37383djdjjd}{\cal F}_- &=& \frac{1}{2}(B_- + g)J_- = \frac{1}{2} d \left ( K_L J 
d X_L + K_C J d\phi + K_T J d\chi \right )~,
\ee
where $B_\pm$ are $(2,0)+(0,2)$-forms with respect to $J_\pm$ and $H=dB_\pm$.  
Using these formulas we can work out explicit expressions for the metric
$g_{\mu\nu}$ and NS-form $H_{\mu\nu\rho}$ and other objects.

For further detail on the local description of generalized \kah geometry and its
relation to supersymmetric sigma models  the reader may consult
\cite{Lindstrom:2005zr, Lindstrom:2007qf, Lindstrom:2007xv, Hull:2008vw}. 

\Section{Appendix: Proof}
\label{appendix2}
\setcounter{equation}{0}

In this appendix we prove that the pure spinors given by (\ref{definitionpurespinor}) 
encode the generalized \kah geometry. Let us start by a reminder of the
relation between the bihermitian and the generalized complex descriptions of
generalized \kah geometry.
\begin{theorem} [{\rm Gualtieri's map \cite{gualtieriPhD}}]
For the generalized \kah geometry the corresponding 
generalized complex structures ${\cal J}_{1,2}$ can be reconstructed from 
the data $(g, J_\pm, B)$
\be
\label{gencompgual}
{\cal J}_{1,2} = 
\frac{1}{2} \left ( \begin{array}{cc}
1 & 0 \\
- B & 1 \end{array} \right )
  \left( 
\begin{array}{cc}
J_+\pm J_- & -(\omega_+^{-1}\mp \omega_-^{-1})\\
(\omega_+\mp\omega_-) & -(J_+^t\pm J_-^t)
\end{array}
\right) 
\left ( \begin{array}{cc}
1 & 0 \\
 B & 1 \end{array} \right )
\ee
 where $H= dB$. 
\end{theorem}

\begin{theorem}
The generalized complex structures ${\cal J}_{1,2}$ correspond to the closed pure
spinors $
\rho_{1,2}$ defined in (\ref{definitionpurespinor}) with $B$ given below in 
(\ref{definitionofB}) such that $H= dB$.
\end{theorem}

Our goal is to show that the pure spinors annihilate the same maximally isotropic
spaces as defined in terms of ${\cal J}_{1,2}$. Using Gualtieri's map we see that an 
eigenvector $X+\eta$ of ${\cal J}_{1}$ with eigenvalue $\pm i$ satisfies
\be\label{oneeigen}
J_+^t (g(X)+\eta + i_{X}B) &=& \mp i(g(X) + \eta+ i_X B)~,\\
\nonumber
J_-^t (g(X)-\eta - i_X B ) &=& \mp i(g(X) - \eta - i_X B)
\ee
and an eigenvector $X+\eta$ of ${\cal J}_2$ with eigenvalue $\pm i$ satisfies
\be\label{twoeigen}
J_+^t (g(X)+\eta + i_X B ) &=& \mp i(g(X) + \eta + i_X B)~,\\
\nonumber
J_-^t (g(X)-\eta - i_X B ) &=& \pm i(g(X) - \eta - i_X B)~.
\ee
The generalized complex structures ${\cal J}_{1,2}$ give rise to the decomposition 
$(T \oplus T^*) \otimes \mathbb{C} = L_1^+ \oplus L_1^- \oplus \bar{L} _1^+
\oplus \bar{L}_1^-$, where $L_1^+$ is $+i$-eigenbundle 
for both ${\cal J}_{1,2}$ and $L_1^-$ is $+i$-eigenbundle for ${\cal J}_1$ and $-i$-
eigenbundle for ${\cal J}_2$. Using (\ref{oneeigen}) and (\ref{twoeigen}) and the fact
that the metric is bihermitian we see that these sub-bundles are defined as follows
$$ L_1^+ = \{ X + g(X) - i_X B~ | ~X \in T_+^{(1,0)}\} = 
  \{ X - i_X ( B + i \omega_+) ~| ~X \in T_+^{(1,0)}\} ~,$$
$$ L_1^- = \{ X - g(X) - i_X B~ | ~X \in T_-^{(1,0)}\} =
  \{ X - i_X ( B - i \omega_-) ~| ~X   \in T_-^{(1,0)}\}~, $$
$$ \bar{L}_1^+ = \{ X + g(X) - i_X B~ | ~X \in T_+^{(0,1)}\} =
  \{ X - i_X ( B - i \omega_+) ~| ~X \in T_+^{(0,1)}\} ~,$$
$$ \bar{L}_1^- = \{ X - g(X) - i_X B~ | ~X \in T_-^{(0,1)}\} =
  \{ X - i_X ( B + i \omega_-) ~| ~X \in T_-^{(0,1)}\}~. $$  
The pure spinor $\rho_1$ should annihilate 
$\bar{L}_1 = \bar{L}_1^+ \oplus \bar{L}_1^-$, i.e. for any $X + \eta \in \bar{L}_1$
\be
\iota_X \rho_1 + \eta \wedge \rho_1 = 0
\ee
and the pure spinor $\rho_2$ should annihilate
$L_1^- \oplus \bar{L}_1^+$, i.e. for any $X + \eta \in   L_1^- \oplus \bar{L}_1^+$
\be
\iota_X \rho_2 + \eta \wedge \rho_2 = 0~.
\ee
Using the concrete form of $\rho_{1,2}$ and the different eigenbundles  we get the 
following relations:
\be
 \label{A111} X \in T^{(0,1)}_+~,~~~~i_X N_1=0~,~~~i_X (R_1 + i S_1 + i \omega_+ - B) 
 \wedge N_1=0~,\\
 \label{A112} X \in T^{(0,1)}_-~,~~~~i_X N_1=0~,~~~i_X (R_1 + i S_1 - i \omega_- - B) 
 \wedge N_1=0~,\\
 \label{A113} X \in T^{(0,1)}_+~,~~~~i_X N_2=0~,~~~i_X (R_2 + i S_2 + i \omega_+ - B) 
 \wedge N_2=0~,\\
  \label{A114} X \in T^{(1,0)}_-~,~~~~i_X N_2=0~,~~~i_X (R_2 + i S_2 + i \omega_- - B) 
  \wedge N_2=0~,
\ee
 
From appendix \ref{appendix1} we have the relations (\ref{relaGKtwosymprl}) and
(\ref{37383djdjjd}), 
\be
\label{fromFplus} 
  B_+  J_+ - \omega_+ &= & d (-K_R J dX_R - K_C Jd\phi + K_T J d\chi) ~,\\
 \label{fromFminus} 
  B_- J_- + \omega_- &= & d(K_L J dX_L + K_C J d\phi + K_T J d\chi )~,
\ee
where $H= dB_+ = dB_-$ and $B_\pm$ is a $(2,0) + (0,2)$ form with respect to
$J_\pm$ respectively. By direct computation one may show that
\be
B_+ - B_- = d(K_T d\chi - K_C d\phi)
\ee
and we can define $B$ as follows
\be
 \label{definitionofB}
  B= B_+ + d(K_C d\phi) = B_- + d(K_T d\chi)~.
\ee
Exactly this $B$ appears in the statement of our theorem. Using these relations it is 
straightforward to prove the relations (\ref{A111})-(\ref{A114}). We will illustrate the
proof for (\ref{A111}) and the rest of the relations (\ref{A112})-(\ref{A114}) are
proven in the same way. For (\ref{A111}), the relation $i_X N_1=0$ is obviously
satisfied since $N_1$ is holomorphic. Using the explicit definition of $R_1$, $S_1$
and the properties (\ref{fromFplus}) and (\ref{definitionofB}) we get
\be
  i_X (R_1 + i S_1 + i \omega_+ - B) \wedge N_1= \nn \\
   i_X \left[ - d(K_L (1- iJ) dX_L) - d(K_C (1-iJ)d\phi ) \right] \wedge N_1 ~,\nn
\ee
where we also used the fact that $B_+$ is $(2,0)+(0,2)$-form with respect to
$J_+$ and $X \in T^{(0,1)}_+$. 

Finally $d(K_L (1- iJ) dX_L)$ is a $(2,0)$-form\footnote{For the explanation 
of this fact see \cite{Lindstrom:2005zr}. Otherwise it 
can be checked explicitly using the form of the complex structures.} with
respect to $J_+$
and thus $i_X (d(K_L (1- iJ) dX_L) =0$ and 
$d(K_C (1-iJ)d\phi ) \wedge N_1=0$ since $N_1$ already contains all of the
$d\boldsymbol{\phi}$. That ends the proof of (\ref{A111}). After similarly
proving (\ref{A112})-(\ref{A114}) we conclude that the pure spinors encode the
correct information about the generalized \kah geometry. 
    
Moreover we can check that the pure spinors $\rho_{1,2}$ obey the correct relation
with respect to the Mukai pairing, $(\rho_1,\rho_2)=(\rho_1,\bar{\rho}_2) = 0$. 
Then the only remaining nontrivial condition comes from 
$(\rho_1,\bar{\rho}_1) = \alpha  (\rho_2,\bar{\rho}_2) \neq 0$ which we study in
detail in section \ref{G-MA}. The conditions
$(\rho_1,\rho_2)=(\rho_1,\bar{\rho}_2) = 0$ can be checked by direct calculation. 
Calculating $(\rho_1,\rho_2)$ we get
\be
  \label{jdjd3388jss}
  N_1\wedge N_2\wedge (R_1-R_2+i(S_1 - S_2))^{(2 d_s+
  \left|\frac{d_c+d_t}{2}\right|)}~,
\ee
where $|\frac{d_c+d_t}{2}|$ indicates the integer part and
\be
R_1 - R_2 +i(S_1-S_2) = - d(K_C(1-iJ)d\phi + K_T(1-iJ)d\chi + 2K_L(1-iJ)dX_L)~.
\ee
Because of the $N_1\wedge N_2$ prefactor in (\ref{jdjd3388jss}), terms containing
$d\boldsymbol{\phi}$ or $d\boldsymbol{\chi}$ in the parenthesis are projected out
and $(\rho_1,\rho_2)$ becomes equal to
\be
N_1\wedge N_2\wedge(2d(K_L(1-iJ)dX_L))^{(2d_s+\left|\frac{d_c+d_t}{2}\right|)}~.
\ee

The  two-form $d(K_L(1-iJ)dX_L)$ is of $(2,0)$-type. with respect to $J_+$ 
and the maximum nonzero power of this term is $d_s$. Thus we conclude that  $
(\rho_1,\rho_2)=0$. 
Analogously we can prove $(\rho_1,\bar{\rho}_2)=0$. The Mukai pairing
$(\rho_1,\bar{\rho}_2)$ is
\be
N_1\wedge\bar{N}_2\wedge(R_1-R_2+i(S_1+S_2))^{(2d_s+ 
  \left|\frac{d_c+d_t}{2}\right|)}~,
\ee
where
\be
R_1-R_2+i(S_1+S_2) = d(K_C(1-iJ)d\phi + K_T(1+iJ)d\chi + 2K_R(1-iJ)dX_R)~.
\ee
Thus $(\rho_1,\bar{\rho}_2)$ becomes equal to
\be
\label{jssk333}
 N_1\wedge \bar{N}_2 \wedge (2d(K_R(1-iJ)dX_R)^{(2d_s+ 
 \left|\frac{d_c+d_t}{2})\right|)}~,
\ee
and here two-form $d(K_R(1-iJ)dX_R)$ is of
$(2,0)$-type  with respect to  $J_-$ and therefore (\ref{jssk333}) vanishes  identically.

\end{document}